\documentclass[11pt]{article}
\usepackage{moriond}

\def\Journal#1#2#3#4{{#1} {\bf #2}, #3 (#4)}

\def\NPB{{\em Nucl. Phys.} B}
\def\PLB{{\em Phys. Lett.}  B}
\def\PRL{\em Phys. Rev. Lett.}
\def\PRD{{\em Phys. Rev.} D}
\def\PRA{{\em Phys. Rev.} A}

\def\PRSA{{\em Proc. Roy. Soc.} A}
\def\NPB{{\em Nucl. Phys.} B}
\def\EPL{\em Europhys. Lett.}
\def\PREP{\em Phys. Rep.}
\def\EPJC{{\em Eur. Phys. J.} C}
\def\EPJP{{\em Eur. Phys. J.} Plus}
\def\APP{\em Astropart. Phys.}
\def\RPP{\em Rep. Prog. Phys.}
\def\RAM{\em Radiat. Meas.}

\def\be{\begin{equation}}
\def\ee{\end{equation}}
\def\bea{\begin{eqnarray}}
\def\eea{\end{eqnarray}}

\newcommand{\Photo}{}

\begin{document}
\vspace*{4cm}
\title{Magnetic monopoles at the LHC and in the Cosmos}

\author{P. Mermod}

\address{D\'epartement de Physique Nucl\'eaire et Corpusculaire, University of Geneva, \\ 1211 Geneva, Switzerland}

\maketitle\abstracts{
The magnetic monopole was postulated in 1931 by Dirac to explain electric charge quantisation. Searches for pair-produced monopoles are performed at accelerator facilities whenever a new energy regime is made available. In addition, monopoles with masses too high to be accessible at colliders would still have been produced in the early Universe and such relics can be searched for either in flight or trapped in matter. Here we discuss recent results and future prospects for direct monopole detection at the LHC and in bulk matter searches, with emphasis on the complementarity between the various techniques. Significant improvements of the results from the ATLAS experiment are expected with the development of new triggers. Dedicated LHC experiments will allow to probe wider ranges of monopole charges and masses: the MoEDAL experiment using both nuclear-track detectors and absorbing arrays, and searches for trapped monopoles in accelerator material. Finally, it is highlighted how the first search for monopoles trapped in polar volcanic rocks allows to set new constraints on the abundance of monopoles bound to matter in the Earth's interior and by extension in the primary material that formed the Solar System. 
}

\section{Introduction}

One example of a well-motivated particle whose discovery would be of great significance to modern physics is the magnetic monopole -- a particle carrying a positive or negative magnetic charge in analogy to the electric charge carried by an electron. The monopole was postulated by Dirac in 1931 to explain electric charge quantisation~\cite{Dirac1931} and was shown by 't Hooft and Polyakov in 1974 to be an essential ingredient in grand-unification theories~\cite{Hooft1974}. It follows from Dirac's argument that monopoles should carry a magnetic charge equal to a multiple of a fundamental unit referred to as the Dirac charge. The Dirac charge is equivalent to 68.5 electron charges in terms of ionisation energy loss at high velocities; monopoles are thus expected to manifest themselves as very highly ionising particles. 

Searches for direct monopole production have been performed each time an accelerator of a new type or surpassing power has been operated, including the CERN Large Electron-Positron (LEP) collider and the Tevatron proton-antiproton collider at Fermilab, where isolated monopoles carrying one Dirac charge were excluded up to masses of the order of $400$ GeV~\cite{SMPs,Bertani:1990tq,OPALdirect,CDFdirect}. Much higher masses (up to 4000 GeV) can be probed at the LHC. It has been pointed out that, to be effective, LHC searches for monopoles should use several complementary techniques~\cite{SMPpheno11}, including the ATLAS general-purpose experiment, the dedicated MoEDAL experiment, and trapping experiments. As discussed below, such a wide programme of searches has already started.  

There are no theoretical constraints on the mass a monopole should carry. The monopole mass can be as high as 10$^{16}$ GeV in grand unification theories. Monopoles with masses above the LHC collision energy could only be produced in high-energy cosmic events, e.g. during the inflation phase of the early Universe~\cite{Lyth99}. Monopoles would be stable and could still be around us today. Even though there are presently no adequate models that describe to which extent relic monopoles would be present in cosmic rays or have accumulated inside astronomical bodies, fluxes and abundances can be constrained by experiments. Monopoles in flight have been sought with array detectors. These set tight constraints on the flux of cosmic monopoles incident on Earth~\cite{Indu91b,MACRO2002,SLIM2008,BAIKAL2008,RICE2008,AMANDA2010,ANITA2011,ANTARES2012} (only the most recent results are given here; see~\cite{PDG2012} for a complete list). Trapped monopoles have been sought in hundreds of kilograms of samples from the Earth's crust~\cite{Kovalik86,Longo95}, in rocks from the Moon's surface~\cite{Moon73}, and in meteorites~\cite{Longo95}. One of the great advantages with meteoritic samples is the possibility that they contain so-called ``stellar'' monopoles, i.e., monopoles which were already trapped in the matter that formed the Solar System. Stellar monopoles would have sunk to the cores of large astronomical bodies such as the Earth and the Moon while they were still molten. However, the Earth's magnetic field is expected to cause the migration of such monopoles along the Earth's magnetic axis. A novel search for stellar monopoles performed recently in polar volcanic rock samples~\cite{SQUIDRocks13} is presented below. 

\section{Monopole searches at the LHC}

Three complementary techniques have been proposed to probe monopole production at the LHC in the most thorough possible way~\cite{SMPpheno11}: in-flight detection in general-purpose detectors, in-flight detection using nuclear-track detectors, and the detection of stopped monopoles in matter with the induction technique. Only direct monopole detection is considered here. A recent discussion of indirect signatures at the LHC can be found e.g. in Ref.~\cite{diphotonLHCpheno12}.

\subsection{Monopoles in the ATLAS detector}

The ATLAS collaboration performed pioneering highly-ionising particle searches with 7 TeV proton-proton collision data, which already largely surpass the LEP and Tevatron results~\cite{QballATLAS10,MonoATLAS11}. ATLAS plans to extend the search with 8 TeV and 14 TeV data. 

A monopole event in the ATLAS detector would be rather striking, with a large number of high-threshold hits in the transition radiation tracker (TRT) and a very localised energy deposition in the electromagnetic (EM) calorimeter. These two independent features are powerful discriminants against backgrounds~\cite{QballATLAS10,MonoATLAS11}. This is assuming, however, that the monopole penetrates all the way to the EM calorimeter. Current level-1 triggers used for collision runs in ATLAS are only sensitive to monopoles which do not range out before entering the EM calorimeter because they rely on either a calorimeter energy deposition or a track in the muon system. For a monopole with the Dirac charge, this corresponds to kinetic energies above 200 GeV~\cite{SMPpheno11}. Moreover, standard electron and photon triggers at level-2 require energy deposition in the second EM calorimeter layer and have transverse energy thresholds exceeding 60 GeV for the data taken in 2011$-$2012. These two conditions increase the minimum kinetic energy required for a Dirac monopole to 500 GeV~\cite{MonoATLAS11}, severely limiting the search acceptance. Therefore, a dedicated level-2 trigger algorithm was developed. This new trigger does not require calorimeter energy deposition beyond level-1 (without second layer requirement and with an 18 GeV threshold) and discriminates on the TRT hit energy to keep the rate under control. This reduces the minimum monopole energy for a fully efficient trigger to 300 GeV. This trigger was enabled in September 2012 and collected 7.1~fb$^{-1}$ of 8 TeV collision data. The analysis of these data will result in the strongest LHC constraints on the production of long-lived massive particles with electric charge up to 100 times the electron charge or magnetic charge up to twice the Dirac charge. 

\subsection{Trapped monopoles in accelerator material}

Magnetic monopoles which would stop in material around collision points without being seen by sensitive detectors would not necessarily be lost. Stopped monopoles are expected to bind to nuclei~\cite{Milton2006} and thus remain trapped in matter. After the run is finished, the material can be taken apart and probed for monopoles with the so-called induction technique, seeking for a persistent current after passage of the samples through a superconducting coil. This kind of search was performed with beam pipes and detector material at the Tevatron and HERA colliders~\cite{TEVATRONSQUID2000,TEVATRONSQUID2004,HERASQUID}. Monopoles need only to traverse vacuum before they reach the beam pipe. Beam-pipe searches would therefore allow to probe magnetic charge ranges which extend well beyond the reach of other detectors~\cite{SMPpheno11}. 

Superconducting magnetometer tests recently performed at the Laboratory of Natural Magnetism at ETH Zurich with calibration devices and obsolete LHC accelerator parts near the CMS interaction region demonstrated that such a search is indeed feasible~\cite{SQUIDtests}. One attractive possibility is to search for monopoles trapped in the central sections of the beryllium beam pipes of the ATLAS and CMS experiments. These beam pipes were exposed to the products of high-energy collisions between 2010 and the end of 2012. They are being replaced and will become available for processing into suitable samples as soon as the performance of the new beam pipes is demonstrated, which is foreseen to happen in $2016-2017$. This is the only experiment which can efficiently probe monopoles with magnetic charge higher than nine times the Dirac charge~\cite{SMPpheno11}. 

\subsection{Monopoles in the MoEDAL detector}

The MoEDAL experiment~\cite{MoEDAL} consists of detector arrays made of thin plastic foils called nuclear track detectors~\cite{NTDs2009} to be exposed to collision products around the LHCb interaction point. After exposure, the nuclear track detectors are removed, etched by a chemical process, and scanned to search for the typical fingerprint a highly-ionising particle would leave behind while traversing the plastic. There are several advantages with such a method: the amount of material between the interaction point and the detectors is typically less than in a general-purpose detector, providing sensitivity to higher charges; the detectors are specifically designed to detect highly-ionising particles in a robust manner and can be calibrated with ion beams; and there are no timing constraints, providing sensitivity to low particle velocities and thus high masses. 

The recent addition of the Magnetic Monopole Trapper (MMT) enhances the sensitivity and redundancy of the MoEDAL experiment. The MMT consists of an absorbing array made of aluminum. Monopoles which would range out and stop within the array would be detectable with the induction technique. The analysis of magnetometer data is relatively fast and straight-forward compared to the complex ATLAS detector or the MoEDAL nuclear track detectors. Due to its bulk and weight, the MMT subdetector can only cover a limited solid angle and has therefore a lower acceptance than the rest of MoEDAL; however, the MMT has the three attractive advantages of simplicity, speed of result delivery, and complementarity. A test array comprising 198 aluminium rods of 60 cm length and 2.5 cm diameter was deployed in September 2012 and is expected to yield its first results in 2013. With limited sensitivity, this first MMT run allows to probe monopoles carrying a magnetic charge larger than the Dirac charge (up to 4 times the Dirac charge) for the first time at the LHC. The full array to be deployed in 2014 will be thicker and will cover more solid angle. This experiment will result in the first monopole search in 14 TeV proton-proton collisions and has the potential to procure a robust and independent cross-check of a discovery as well as a unique measurement of the magnetic properties of a monopole. 

\section{Primordial monopole searches}

Recent searches for monopoles trapped in bulk matter have focused on stellar monopoles, which were missed by previous searches which used samples from the crusts of the Earth and the Moon. Accessible samples which could contain stellar monopoles include meteorites and polar volcanic rocks. 

\subsection{Monopoles in meteorites}

The most extensive meteorite search to date was performed in 1995~\cite{Longo95}. It sets a limit on the monopole density in meteoritic material of less than $2.1\cdot 10^{-5}$/gram at 90\% confidence level. The study analysed 112 kg of meteorites, among which $\sim 100$ kg are chondrites and can thus be assumed to consist of undifferentiated material from the primary solar nebula. 

\subsection{Monopoles in polar volcanic rocks}

Stellar monopoles inside the Earth would tend to accumulate at a point along the magnetic axis where the downwards gravitational force is equal to the upwards force exerted by the Earth's magnetic field. An equilibrium above the core-mantle boundary for a monopole with the Dirac charge corresponds to the condition that the mass should be less than $4\cdot 10^{14}$ GeV. Assuming a binding of monopoles to nuclei, from this point, the solid mantle convection would be expected to slowly bring up monopoles to the surface. Over geologic time, monopoles would thus accumulate in the mantle beneath the geomagnetic poles for a wide range of masses and charges. 

These considerations motivated a novel search which probed 23.4 kg of polar volcanic rock samples~\cite{SQUIDRocks13} with the superconducting magnetometer in Zurich. This represents $4-5$ times less material than used in the meteorite search. However, for monopole mass and charge satisfying the criterion for a position above the core-mantle boundary, this difference is compensated for by an increase in monopole concentration of roughly a factor 6 in polar mantle-derived rocks. It results that, in a simple model, a limit of $1.6\cdot 10^{-5}$/gram can be set in the matter averaged over the whole Earth. This is slightly better than the limit from the meteorite search.

\section{Conclusions}

A wide programme of searches for magnetic monopoles at the LHC has started. The production of monopoles of any charge and mass can be probed efficiently by combining the use of special triggers in the ATLAS general-purpose detector, the analysis of obsolete accelerator material, and the use of dedicated MoEDAL array detectors. 

The polar volcanic rock search provides a novel probe of stellar monopoles in the Solar System. One can think of two ways in which these results could be further improved in the future: by analysing large ($>100$ kg) amounts of meteorites and polar rocks with a high-efficiency magnetometer, or by gaining access to new types of samples such as asteroid and comet fragments. 

\section*{Acknowledgments}

This work was supported by a fellowship from the Swiss National Science Foundation and a grant from the Ernst and Lucie Schmidheiny Foundation.

\end{document}